\documentclass[3p,twocolumn,sort&compress, fleqn]{elsarticle}
\usepackage{amsmath, bm}
\usepackage{amssymb}
\usepackage{amsmath}
\usepackage{braket}
\journal{Computer Physics Communications}

\def\Hloc{{H}_\mathrm{loc}}
\def\tauR{\tau_\mathrm{R}}
\def\tauL{\tau_\mathrm{L}}
\def\G1{{G^{(1)}}}

\newcommand{\ttau}{\ensuremath{\tilde{\tau}}}

\usepackage{natbib}
\usepackage{hyperref}
\hypersetup{
        colorlinks=true,
}

\long\def\beginmypgfpdfnamed#1#2\endmypgfpdfnamed{\includegraphics{#1}}

\graphicspath{{./pyfig/}}

\begin{document}

\begin{frontmatter}

\title{
	Continuous-time hybridization expansion quantum impurity solver for multi-orbital systems with complex hybridizations
}

\author[saitama]{Hiroshi Shinaoka} \ead{shinaoka@mail.saitama-u.ac.jp}
\author[michigan]{Emanuel Gull}
\author[fribourg]{Philipp Werner}

\address[saitama]{Department of Physics, Saitama University, Saitama 338-8570, Japan}
\address[michigan]{University of Michigan, Ann Arbor, Michigan 48109, USA}
\address[fribourg]{Department of Physics, University of Fribourg, 1700 Fribourg, Switzerland}

\begin{abstract}
We describe an open-source implementation of the continuous-time hybridization-expansion quantum Monte Carlo method for impurity models with general instantaneous two-body interactions and complex hybridization functions. 
The code is built on an updated version of the core libraries of ALPS (Applications and Libraries for Physics Simulations) [ALPSCore libraries].
\end{abstract}

\begin{keyword}
Quantum impurity problems \sep continuous-time impurity solver  \sep hybridization expansion \sep complex hybridization functions \sep dynamical mean-field theory
\end{keyword}

\end{frontmatter}
{\bf PROGRAM SUMMARY}

\begin{small}
\noindent
{\em Program Title:} ALPSCore CT-HYB \\
{\em Journal Reference:}                                      \\
{\em Catalogue identifier:}                                   \\
{\em Licensing provisions:} GPLv3\\
{\em Programming language:} \verb*#C++#, MPI for parallelization. \\
{\em Computer:} PC, HPC cluster \\ 
{\em Operating system:} Any, tested on Linux and Mac OS X\\ 
{\em RAM:} 100 MB - 1 GB.\\ 
{\em Number of processors used:} 1 - 2000.\\ 
{\em Keywords:} impurity solver, CT-HYB\\ 
{\em Classification:} 4.4  \\ 
{\em External routines/libraries:}  ALPSCore libraries, Eigen3, Boost.\\ 
{\em Nature of problem:} Quantum impurity problem\\
{\em Solution method:} Continuous-time hybridization-expansion quantum Monte Carlo\\
{\em Running time:} 1 min -- 8 h (strongly depends on the problem to solve)\\
\end{small}

\section{Introduction}
In condensed matter physics, dynamical mean-field theory (DMFT)~\cite{Georges:1996un} is a widely used tool for the study of strongly correlated electron
 systems.
In a DMFT calculation, a correlated lattice model is mapped to an impurity problem whose bath degrees of freedom are self-consistently determined. 
Although the approximation was originally proposed for the single-band Hubbard model, the DMFT formalism can be extended to multi-orbital systems and cluster-type impurities~\cite{Maier:2005et}.
Furthermore, DMFT can be combined with density functional theory based ab-initio calculations, to describe strongly correlated materials such as transition metal oxides~\cite{Kotliar:2006fl}. 
In such realistic applications, one may have to treat local Coulomb interactions with non-density-density terms.
Furthermore, in simulations of $4d$ and $5d$ transition metal oxides,
spin-orbit coupling gives rise to complex hybridization functions~\cite{Shinaoka:2015vma}.
In the presence of multiple local degrees of freedom or complex hybridization functions, the solution of the quantum impurity problem becomes a numerically costly task.
For the DMFT self-consistency loop, we only have to compute single-particle quantities such as the self-energy.
However, it may be desirable to compute also 
higher-order correlation functions, to get access to spin-orbital susceptibilities, or in the context of diagrammatic extensions of the DMFT formalism~\cite{Toschi:2007fq,Rubtsov:2008cs}.

Continuous-time Monte Carlo is a general framework to solve a quantum models in a numerically exact way by sampling a series expansion of the partition function.
There are two complementary algorithms for quantum impurity problems with general interactions, called CT-INT (continuous-time interaction expansion) and CT-HYB (continuous-time bybridization expansion). The former is based on the expansion of the partition function with respect to the local interaction~\cite{Rubtsov:2005iwa}, while the latter is based on an expansion in the hybridization between the bath and the impurity~\cite{Werner:2006ko}.
CT-HYB is particularly efficient in the strongly correlated regime~\cite{Gull:2007gf}, and it is widely used for both model and material simulations.

The available open-source implementations of CT-HYB~\cite{Hafermann:2013il,Seth:2015uq,Huang:2015cc} support neither complex hybridisation functions nor the measurement of the two-particle Green's function for general multi-orbital models.
In this paper, we describe a state-of-the-art implementation of CT-HYB for multiple orbitals and complex hybridization functions. 
The code provides measurements of various observables and correlation functions relevant for DMFT calculations, including the single-particle Green's function, density-density correlations, and four-time/two-time two-particle Green's functions.
The measurements are performed by worm sampling~\cite{Anonymous:HpLFkQay,Gull:2011jda,Gunacker:2015fj,Gunacker:2016vw}, to avoid ergodicity problems.

The remainder of this paper is organized as follows.
In Sec.~2, we introduce a general quantum impurity model.
In Sec.~3, we describe CT-HYB and worm sampling.
Section~4 describes the evaluation of the trace over local degrees of freedom, 
while 
Sec.~5 explains the possibility to change the single-particle basis. 
The installation and usage is detailed in Sec.~6, and
Sec.~7 provides some examples of simulation results for a three-orbital model.
Finally, we summarize the paper in Sec.~8.

\section{General impurity model}

We consider a general multi-orbital impurity model defined by the Hamiltonian
\begin{eqnarray}
	\mathcal{H}_\text{imp} &=& H_\mathrm{loc} + \mathcal{H}_\mathrm{bath} + \mathcal{H}_\mathrm{hyb},
\end{eqnarray}
where
\begin{eqnarray}
	\mathcal{H}_\mathrm{loc} &=& \sum_{ab} t_{ab} c^\dagger_a c_b + \sum_{abcd} V_{abcd}c_a^\dagger c_b^\dagger c_c c_d,\\
	\mathcal{H}_\mathrm{bath} &=& \sum_{\alpha} \epsilon_\alpha d^\dagger_\alpha d_\alpha,\\
	\mathcal{H}_\mathrm{hyb} &=& \sum_{\alpha,b} V_{\alpha b} d^\dagger_\alpha c_b + \mathrm{h.c.}
\end{eqnarray}
The indices $a$ and $b$ denote the internal degrees of freedom of the impurity, which we call \textit{flavors}, e.g., a composite index of spin and orbital,
while $\alpha$ denotes those of the bath.
The chemical potential is absorbed into $t_{ab}$.

For the Monte Carlo simulation, we switch to an action formulation and trace out the bath degrees of freedom. In the action
\begin{align}
\mathcal{S}_\text{imp}=&\int_0^\beta d\tau \mathcal{H}_\text{loc}(\tau)\nonumber\\
&+ \int_0^\beta d\tau d\tau' \sum_{ab} c^\dagger_a(\tau)\Delta_{ab}(\tau-\tau^\prime)c_b(\tau^\prime) 
\end{align}
all relevant local information of $\mathcal{H}_\mathrm{bath}$ and $\mathcal{H}_\mathrm{hyb}$
is encoded in the hybridization function defined by
\begin{eqnarray}
	\Delta_{ab}(i\omega_n) &=& \sum_{\alpha} \frac{V_{a\alpha}^* V_{\alpha b}}{i\omega_n - \epsilon_\alpha}.
\end{eqnarray}
Here, $\beta$ is the inverse temperature and $\omega_n=(2n+1)\pi/\beta$ a Matsubara frequency. 
The Fourier transformation to imaginary time can be written as 
\begin{align}
\Delta_{ab}(\tau)  
         &= \frac{1}{\beta} \sum_{n=-\infty}^\infty e^{-\mathrm{i}\omega_n \tau} \Delta_{ab}(\mathrm{i}\omega_n)\nonumber \\
	&= \frac{1}{\beta} \sum_{n=0}^\infty e^{-\mathrm{i}\omega_n \tau}
	\left\{\Delta_{ab}(\mathrm{i}\omega_n) + \Delta_{ba}^*(\mathrm{i}\omega_n)\right\}.
\end{align}
The hybridization function is fermionic and hence $\beta$-antiperiodic, $\Delta_{ab}(\tau+\beta) = -\Delta_{ab}(\tau)$,
with discontinuities at $\tau = n \beta$ ($n$ integer).
For the simulations, we define $\Delta(\tau)$ on the interval [0, $\beta$] in a continuous fashion.
In the original literature \cite{Werner:2006ko,Werner:2006iz}, there is a alternative notation for the hybridization function, $F_{ab}$,
which is related to $\Delta_{ab}$ by
\begin{eqnarray}
	F_{ab}(\tau) &=& -\Delta_{ba}(\beta-\tau)~(0 \le \tau \le \beta).
\end{eqnarray}
Our impurity solver takes $\Delta(\tau)$ as an input
in addition to the transfer matrix $t_{ab}$ and the Coulomb tensor $U_{abcd}$.
Note that the hybridization function can have diagonal and offdiagonal components, and that their values may be complex. The latter property is for example essential for the simulation of models with spin-orbit coupling~\cite{Shinaoka:2015vma}. The different components must satisfy the relation $\Delta_{ab}(\tau) = \Delta^*_{ba}(\tau)$.

\section{Hybridization expansion and worm sampling}
\subsection{Expansion of the partition function}
For a given inverse temperature $\beta$,
the partition function $Z$ of the impurity model is expanded in powers of the hybridization as
\begin{align}
	& Z  \equiv \braket{e^{-\beta \mathcal{H}} }\nonumber\\
	&\propto  \sum_{n=0}^\infty \frac{1}{n!^2}\sum_{a_1,\cdots,a_n}\sum_{a^\prime_1,\cdots,a^\prime_n}
	  \int_0^\beta \mathrm{d} \tau_1 \mathrm{d} \tau_1^\prime \cdots \int_0^\beta \mathrm{d} \tau_n \mathrm{d} \tau_n^\prime \nonumber\\
	& \times \mathrm{Tr_{loc}}\left[ e^{-\beta\mathcal{H}_\mathrm{loc}} T_\tau c_{a_1}(\tau_1) c^\dagger_{a_1^\prime}(\tau_1^\prime) \cdots c_{a_n}(\tau_n) c^\dagger_{a_n^\prime}(\tau_n^\prime)\right]\nonumber\\
	& \times \mathrm{det} \boldsymbol{M}^{-1},
	\label{eq:z-cthyb}
\end{align}
where $n$ is the order of the expansion, $c_a^\dagger(\tau) = e^{\tau \Hloc} c_a^\dagger e^{-\tau \Hloc}$ and $c_a(\tau) = e^{\tau \Hloc} c_a e^{-\tau \Hloc}$.
The matrix elements of $(\boldsymbol{M}^{-1})_{ij}$ are given by the hybridization function, 
\begin{align}
 (\boldsymbol{M}^{-1})_{ij} &= \Delta_{a_i^\prime,a_j}(\tau_i^\prime - \tau_j).\label{eq:M}
\end{align}
Equation~(\ref{eq:z-cthyb}) can be recast into 
\begin{align}
Z 
&\propto  \sum_{n=0}^\infty \sum_{a_1,\cdots,a_n}\sum_{a^\prime_1,\cdots,a^\prime_n}\nonumber \\
&  \int_0^{\tau_2} \mathrm{d} \tau_1 \int_0^{\tau_2^\prime}\mathrm{d} \tau_1^\prime \cdots \int_0^\beta \mathrm{d} \tau_n \int_0^\beta \mathrm{d} \tau_n^\prime \nonumber\\
& \times \mathrm{Tr}_\mathrm{loc}\Big[
 e^{-(\beta-\ttau_{2n}) \Hloc} O_{2n} \cdots e^{-(\ttau_2-\ttau_1) \Hloc} O_1 e^{-\ttau_1 \Hloc} 
\Big]\nonumber\\
& \times (-1)^{P_\mathrm{trace}} \times \mathrm{det} \boldsymbol{M}^{-1}
,\label{eq:z-cthyb-tordered}
\end{align}
where 
$0\le \tau_1<\tau_2<\ldots <\tau_n < \beta$, $0\le \tau'_1<\tau'_2<\ldots <\tau'_n<\beta$, and 
\{$O_1$, $\cdots$, $O_{2n}$\} is a time-ordered set of the impurity creation and annihilation operators.
 $P_\mathrm{trace}$ is the permutation of
time ordering from \{$c_{a_1}(\tau_1), c^\dagger_{a_1^\prime}(\tau_1^\prime), \cdots , c_{a_n}(\tau_n), c^\dagger_{a_n^\prime}(\tau_n^\prime)$\} 
to \{$O_1$, $\cdots$, $O_{2n}$\}.
The set $\{\ttau_1, \cdots, \ttau_{2n}\}$ is a time-ordered set of $\{\tau_1, \cdots, \tau_n, \tau_1^\prime, \cdots, \tau_n^\prime\}$.
In the Monte Carlo simulations, we perform an importance sampling of $Z$ 
using the configurations $c=\{\tau_1, \tau_1^\prime, \cdots, \tau_n, \tau_n^\prime; a_1, a_1^\prime,\cdots, a_n, a_n^\prime\}$ and the Metropolis algorithm. Since the weight
\begin{align}
w(c)=& \mathrm{Tr}_\mathrm{loc}\Big[ e^{-(\beta-\ttau_{2n}) \Hloc} O_{2n} \cdots\nonumber\\
&  e^{-(\ttau_2-\ttau_1)  \Hloc} O_1 e^{-\ttau_1 \Hloc} \Big](-1)^{P_\mathrm{trace}} \mathrm{det} \boldsymbol{M}^{-1}\label{eq:weight}
\end{align}
can be complex, we consider the absolute value $|w(c)|$ in the sampling.
That is, we sample an unphysical partition function
\begin{align}
  \bar{Z} &\equiv \sum_{c} |w(c)|.
\end{align}
The ratio $\bar{Z}/Z$ appears as a reweighing factor in the estimators of observables.
We measure the fidelity susceptibility in this partition-function space~\cite{2015PhRvX...5c1007W,Wang:2015je}.

\subsection{Worm sampling of the Green's function}
We now briefly describe how to measure the expectation values of correlation functions.
For instance, the single-particle Green's function is defined as
\begin{align}
G_{ij}(\tau - \tau^\prime) &= -\frac{\mathrm{Tr}[T_\tau e^{-S_\mathrm{imp}} c_i(\tau) c^\dagger_j(\tau^\prime)]}{Z},\label{eq:def-G1}
\end{align}
where $\tau, \tau^\prime \in [0,\beta)$.
This quantity is the most important observable for DMFT calculations.

Similarly to Eq.~(\ref{eq:z-cthyb}),
the numerator of Eq.~(\ref{eq:def-G1}) is expanded as 
\begin{align}
& \mathrm{Tr}[T_\tau e^{-S_\mathrm{imp}} c_i(\tau) c^\dagger_j(\tau^\prime)] \nonumber\\
& = Z_\mathrm{B} \sum_{n=0}^\infty \sum_{a_1,\cdots,a_n}\sum_{a^\prime_1,\cdots,a^\prime_n}\nonumber\\
& \int_0^{\tau_2} \mathrm{d} \tau_1 \int_0^{\tau_2^\prime}\mathrm{d} \tau_1^\prime \cdots \int_0^\beta \mathrm{d} \tau_n \int_0^\beta \mathrm{d} \tau_n^\prime \nonumber\\
& \times \mathrm{Tr}_\mathrm{loc}\Big[
e^{-(\beta-\ttau_{2n+2}) \Hloc} O_{2n+2} \cdots \nonumber\\
& \hspace{15mm}\cdots e^{-(\ttau_2-\ttau_1) \Hloc} O_1 e^{-\ttau_1 \Hloc} \Big]\nonumber\\
& \times (-1)^{P_\mathrm{trace}} \mathrm{det} \boldsymbol{M}^{-1},
\end{align}
where \{$O_1$, $\cdots$, $O_{2n+2}$\} is a time-ordered set of the impurity creation and annihilation operators, and those of the Green's function. 
$P_\mathrm{trace}$ is the permutation of
 time ordering from \{$c_{a_1}(\tau_1), c^\dagger_{a_1^\prime}(\tau_1^\prime), \cdots , c_{a_n}(\tau_n), c^\dagger_{a_n^\prime}(\tau_n^\prime), c_i(\tau), c_j^\dagger(\tau^\prime)$\} 
 to \{$O_1$, $\cdots$, $O_{2n+2}$\}.
The set $\{\ttau_1, \cdots, \ttau_{2n+2}\}$ is a time-ordered set of $\{\tau_1, \cdots, \tau_n, \tau_1^\prime, \cdots, \tau_n^\prime, \tau, \tau^\prime\}$.
 
To construct the estimator of Eq.~(\ref{eq:def-G1}),
we enlarge the configuration space as~\cite{Anonymous:HpLFkQay,Gull:2011jda,Gunacker:2015fj}
\begin{align}
C &= C_Z \oplus C_{G^{(1)}},
\end{align}
where a configuration $c_A$ in the Green's function space $C_A$ is defined as
$c_A=\{\tau_1, \tau_1^\prime, \cdots, \tau_n, \tau_n^\prime; a_1, a_1^\prime,\cdots, a_n, a_n^\prime; c_i(\tau), c_j^\dagger(\tau^\prime)\}$.
The contribution of $c_A$ to the expectation value reads
\begin{align}
&w(c_A) = \mathrm{Tr}_\mathrm{loc}\Big[
e^{-(\beta-\ttau_{2n+2}) \Hloc} O_{2n+2} \cdots \nonumber\\
& \cdots e^{-(\ttau_2-\ttau_1) \Hloc} O_1 e^{-\ttau_1 \Hloc} \Big] (-1)^{P_\mathrm{trace}} \mathrm{det} \boldsymbol{M}^{-1},\label{eq:weightA}
\end{align}
up to a normalization factor.
We sample in both configuration spaces according to weight $|w(c)|$ and $|w(c_A)|$, respectively.
In practice, we switch from $C_Z$ to $C_{G^{(1)}}$ by inserting a ``worm" $\{c_i(\tau)$,$c_j^\dagger(\tau^\prime)\}$.
We return back to $C_Z$ by removing a worm.

The total partition function $\bar{W}$ is defined as
\begin{align}
  \bar{W} &= \bar{Z} + \eta \bar{Z}_{G^{(1)}}
\end{align}
with
\begin{align}
  \bar{Z}_{G^{(1)}} &= \int_0^\beta \left|\mathrm{Tr}[T_\tau e^{-S_\mathrm{imp}} c(\tau) c^\dagger(\tau^\prime)]\right|.
\end{align}
The overline means that we take the absolute values of the contributions of diagrams.
The coefficient $\eta~(>0)$ was introduced so that the simulation spends almost the same number of Monte Carlo steps in both spaces. 
It appears as an additional factor in the weight $|w(c_A)|$.
The parameter $\eta$ is adjusted during the thermalization process using the Wang-Landau algorithm~\cite{Wang:2001eb,Wang:2001jk}.
We refer the reader to \ref{sec:wang-landau} for more details.

Considering the anti-periodicity of the Green's function with respect to $\beta$,
the estimator reads
\begin{align}
G(\Delta \tau) &= \frac{\beta^{-1} \eta^{-1} N_{G^{(1)}} \braket{\mathrm{sign}~\tilde{\delta}(\Delta \tau - (\tau-\tau^\prime))}_{\G1}}{N_Z \braket{\mathrm{sign}}_{Z}},\label{eq:G1-it}
\end{align}
for $0 < \Delta \tau < \beta$.
To simplify the notation, we introduced
\begin{align}
  & \tilde{\delta}(\tau) =
  \begin{cases}
   \delta(\tau) & 0 \le \tau < \beta\nonumber \\
   -\delta(\tau + \beta) & -\beta \le \tau < 0\nonumber \\   
  \end{cases}.
\end{align}
The symbols ``$\mathrm{sign}$" in the numerator and the denominator denote $w(c_\G1)/|w(c_\G1)|$ and $w(c_Z)/|w(c_Z)|$, respectively.
$N_Z$ and $N_{\G1}$ are the number of Monte Carlo steps spent in $C_Z$ and $C_{G^{(1)}}$, respectively.
The brackets $\braket{\cdots}_Z$ and $\braket{\cdots}_\G1$ denote the Monte Carlo average in $C_Z$ and $C_\G1$, respectively.
The factor $\beta^{-1}$ comes from the extra degree of freedom $\tau^\prime$ in the sampling in $C_\G1$.

In general, $w(c_Z)/|w(c_Z)|$ is a complex number, but the expectation value $\braket{\mathrm{sign}}_Z$ is real, because the partition function $Z$ is real. 
The importance sampling works efficiently as long as $\braket{\mathrm{sign}}_Z$ is not too small ($\text{sign}\gtrsim 0.1 $).

In practice, instead of using the imaginary-time estimator, Eq.~(\ref{eq:G1-it}), we expand the Green's function in the Legendre polynomials defined on the interval [0, $\beta$]~\cite{Boehnke:2011dd} as
\begin{align}
	G_{ab}(\tau) &= \sum_{l\ge 0}^{N_l} \frac{\sqrt{2l+1}}{\beta} P_l [x(\tau)] G^{ab}_l,\\
	G^{ab}_l &= \sqrt{2l+1} \int_0^\beta d\tau P_l[x(\tau)]) G(\tau),
\end{align}
where $x(\tau) = 2\tau/\beta-1$ and $P_l(x)$ is the $l$-th Legendre polynomial defined on the interval [-1,1].
In the Legendre basis, the estimator reads
\begin{align}
G_l^{ab} &= \frac{\beta^{-1} \eta^{-1} N_{\G1}\sqrt{2l+1} \braket{\mathrm{sign}~\tilde{P}_l(x(\tau-\tau^\prime)}_{\G1}}{N_Z \braket{\mathrm{sign}}_{Z}},\label{eq:G1-legendre}
\end{align}
where
\begin{align}
	\tilde{P}_l[x(\delta \tau)] &= 
	  \begin{cases}
	  	P_l[x(\delta \tau)],~(\delta \tau>0),\\
	  	-P_l[x(\delta \tau+\beta)],~(\delta \tau<0).
	  \end{cases}
\end{align}
The cutoff $N_l$ is a simulation parameter.
Typical values are $N_l=50$ -- $100$.

At this point, it is worth pointing out a practical limitation of this measurement.
In the estimator (\ref{eq:G1-legendre}),
we measure the Green's function only at one time difference and a pair of flavors at each measurement step.
Thus, the sampling by this estimator is less efficient than the conventional measurement by removing hybridization lines from a configuration in $C_Z$ as in the conventional CT-HYB \cite{Werner:2006ko}.

To improve the statistics, 
we generate multi samples from a configuration in $C_\G1$ by reconnecting hybridization lines at each worm measurement step without reevaluating the trace. 
We refer the reader to \ref{sec:meas-G} for more details.

\subsection{Worm measurement of other quantities}
We also provide worm measurements of the following correlation functions, which play an important role in DMFT calculations:
\begin{itemize}
\item Two-particle Green's function
\begin{align}
& G_{abcd}^{(2)}(\tau_1, \tau_2, \tau_3, \tau_4) =\nonumber\\
& \frac{\mathrm{Tr}[T_\tau e^{-S_\mathrm{imp}} c_a(\tau_1) c^\dagger_b(\tau_2) c_c(\tau_3) c^\dagger_d(\tau_4)]}{Z},
\end{align}

\item Equal-time single-particle Green's function~\cite{Gunacker:2016vw}
\begin{align}
& \frac{\mathrm{Tr}[T_\tau e^{-S_\mathrm{imp}} c^\dagger_a(\tau) c_b(\tau)]}{Z},
\end{align}

\item Two-time two-particle Green's function
\begin{align}
& \frac{\mathrm{Tr}[T_\tau e^{-S_\mathrm{imp}} c^\dagger_a(\tau)c_b(\tau) c^\dagger_c(\tau^\prime)c_d(\tau^\prime)]}{Z}.\label{eq:two-time-G2}
\end{align}
\end{itemize}
We measure the two-particle Green's function and the two-time two-particle Green's function in the Legendre basis.

The extended configuration space for the measurement of multiple observable reads
\begin{align}
  \bar{W} &= \bar{Z} + \eta \bar{Z}_\G1 + \eta_{{\cal O}_1} \bar{Z}_{{\cal O}_1} + \cdots + \eta_{{\cal O}_N} \bar{Z}_{{\cal O}_N},
\end{align}
where ${\cal O}_1, \cdots, {\cal O}_N$ are the observables measured by worm sampling.
The coefficient $\eta, \eta_{{\cal O}_1}, \cdots, \eta_{{\cal O}_N}$ are adjusted so that the simulation spends an equal amount of steps in each subspace.

Worm insertion/removal steps result in transitions between $\bar{Z}$ and the worm subspaces $\bar{Z}_{{\cal O}_i}$.
We also perform direct transitions between the worm spaces of different observables such as the equal-time single-particle Green's function and the two-time two-particle Green's function to reduce autocorrelation times.

\section{Evaluation of the trace over the local Hilbert space}
\subsection{Construction of eigenbasis}
We briefly describe how we evaluate Eq.~(\ref{eq:weight}), which is the heart of the CT-HYB code.
The size of the local Hilbert space grows exponentially with the number of orbitals or sites in the impurity.
The computational cost is greatly reduced by partitioning the local Hilbert space into sectors such that $\mathcal{H}_\mathrm{loc}$ is a block diagonal matrix and the block matrices of creation and annihilation operators have only a single nonzero block for each row~\cite{Haule:2007kx}.

After partitioning the Hilbert space into sectors,
we construct an eigenbasis for each sector to rewrite the trace as
\begin{align}
& \mathrm{Tr}_\mathrm{loc}\Big[ e^{-(\beta-\tau_{2n}) \Hloc} O_{2n} \cdots e^{-(\tau_2-\tau_1) \Hloc} O_1 e^{-\tau_1 \Hloc} \Big]\nonumber\\
& =\sum_s \mathrm{Tr}\Big\{ \boldsymbol{E}^{s,s_{2n}}(\beta-\tau_{2n}) \boldsymbol{O}_{2n}^{s_{2n},s_{2n-1}} \cdots \nonumber\\
& \cdots \boldsymbol{E}^{s_1}(\tau_2-\tau_1) \boldsymbol{O}_1^{s_1,s} \boldsymbol{E}^s(\tau_1) \Big\},\label{eq:sectored-trace}
\end{align}
where $\boldsymbol{E}^s(\tau)$ is the diagonal 
matrix $\{e^{-\epsilon_n^s \tau}\}$ ($\epsilon_n^s$ is the $n$-th eigenvalue of the sector $s$).
$\boldsymbol{O}^{s,s^\prime}$ is the matrix representation of an annihilation $d$ or a creation operator $d^\dagger$.
For a given sector $s^\prime$,
there is only one non-zero block matrix $\boldsymbol{O}^{s,s^\prime}$ ($s$ is determined uniquely).

The partitioning of the Hilbert space may be done by exploiting conserved quantum numbers that commute with $\mathcal{H}_\mathrm{loc}$.
Examples include the total electron number as well as 
the special conserved quantities which commute with the Slater-Kanamori Hamiltonian~\cite{Parragh:2012hh}.
Recently, it was shown that it is possible to partition the local Hilbert space without prior knowledge of quantum numbers for CT-HYB~\cite{Parragh:2013us,Seth:2015uq}.
This is done by looking at non-zero elements of the matrix representations of $\mathcal{H}_\mathrm{loc}$ and $c^\dagger_a$ in the occupation basis.
Our code provides the same functionality based on an efficient cluster identification algorithm. 
We refer the readers to \ref{sec:partioning} for more technical details.

\subsection{Sliding-window update}
\begin{figure}
	\centering\includegraphics[width=.4\textwidth,clip]{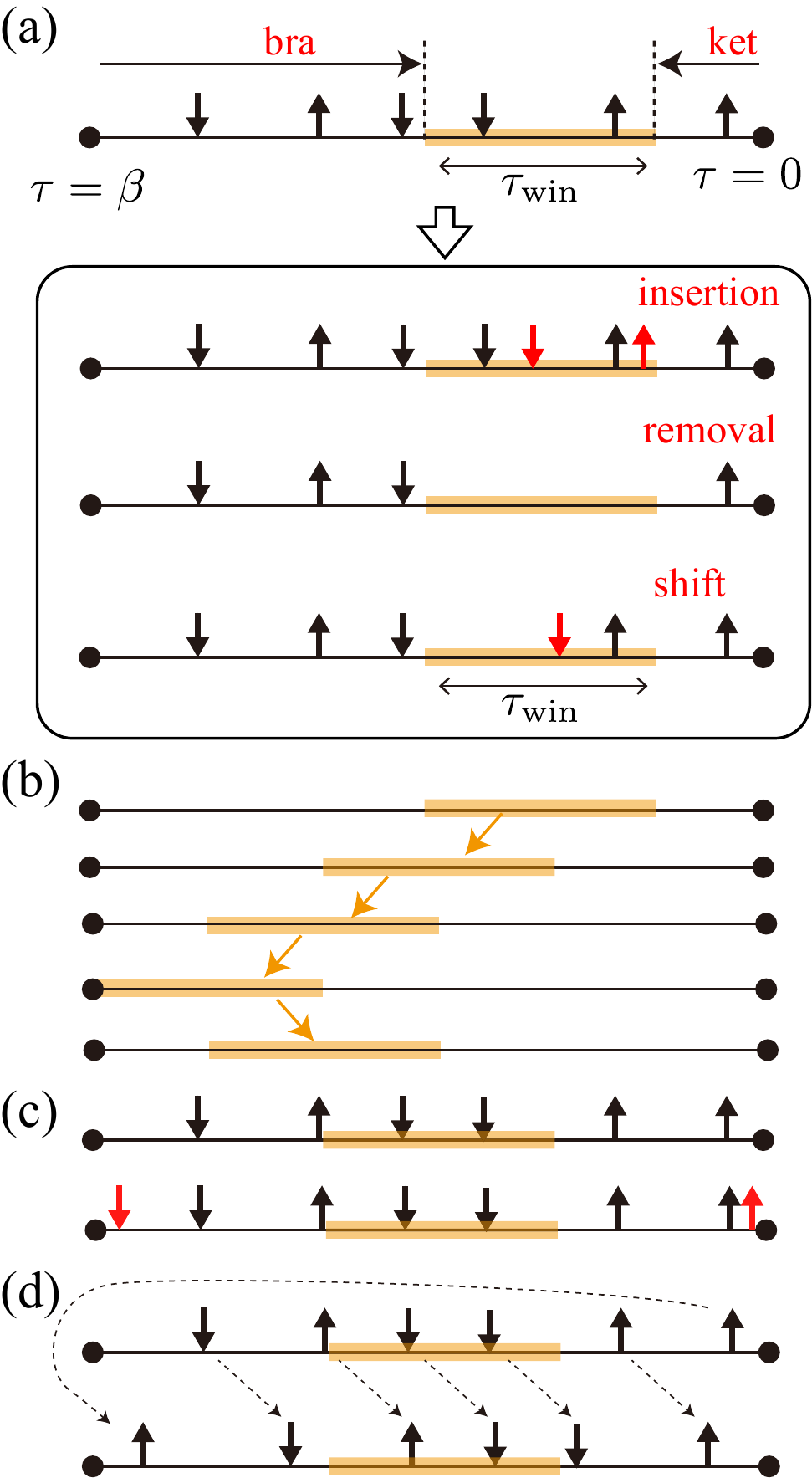}
	\caption{
		(Color online) 
		(a) Elementary updates in the sliding window.
		(b) Sequential sweep of the window on the imaginary time axis.
		(c) Insertion of creation and annihilation operators crossing the boundary at $\beta=0$.
		(d) Global update in which all operators are shifted on the imaginary time axis.
	}
	\label{fig:sliding-window}
\end{figure}
Monte Carlo updates consist of elementary updates such as the insertion/removal of a pair of $c$ and $c^\dagger$.
To compute the acceptance rate of a new configuration,
we evaluate the trace using Eq.~(\ref{eq:sectored-trace}).
This may cost $O(\beta)$ operations because the number of matrices to be multiplied increases linearly with $\beta$.
We reduce the computational cost by using the sliding-window update proposed in Ref.~\cite{Anonymous:JWG_UROR}.
As illustrated in Fig.~\ref{fig:sliding-window}(a),
we define a narrow window in which the updates are performed on the imaginary time axis.
The left and right end points are $\tau_\mathrm{L}$ and $\tau_\mathrm{R}$ ($\tau_\mathrm{L}-\tau_\mathrm{R}=\tau_\mathrm{win}$).
The idea is that we precompute the products of all matrices for $\tau>\tauL$ and $\tau<\tauR$, respectively, and store them in memory.

We define a \textit{ket} as
\begin{align}
 \ket{\tau_\mathrm{R}, s}=& \boldsymbol{R}(\tau_\mathrm{R}, s)\nonumber\\
 \equiv & \boldsymbol{E}^{s_{n_\mathrm{R}+1},s_{n_\mathrm{R}}}(\tau_\mathrm{R}-\tau_{n_\mathrm{R}}) \boldsymbol{O}_{n_\mathrm{R}}^{s_{n_\mathrm{R}},s_{n_\mathrm{R}-1}} \cdots\nonumber\\
 & \cdots \boldsymbol{E}^{s_1}(\tau_2-\tau_1) \boldsymbol{O}_1^{s_1,s} \boldsymbol{E}^s(\tau_1),\label{eq:ket}
\end{align}
where $n_\mathrm{R}$ is the number of operators on the interval ($\tau_\mathrm{R}$,0].
We define a \textit{bra} in a similar way as
\begin{align}
\bra{\tau_\mathrm{L}, s}=& \boldsymbol{L}(\tau_\mathrm{L}, s)\nonumber\\
\equiv &\boldsymbol{E}^{s,s_{2n}}(\beta-\tau_{2n}) \boldsymbol{O}_{2n}^{s_{2n},s_{2n-1}} \cdots\nonumber\\
& 
 \cdots \boldsymbol{O}_{n_\mathrm{L}}^{s_{n_\mathrm{L}},s_{n_\mathrm{L}-1}}
 \boldsymbol{E}^{s_{n_\mathrm{L}-1}}(\tau_{n_\mathrm{L}}-\tau_\mathrm{L}),\label{eq:bra}
\end{align}
where $n_\mathrm{L}$ is the index of the operator with the smallest imaginary time on the interval ($\beta$, $\tau_\mathrm{L}$].
Then, the trace is rewritten as
\begin{align}
 & \sum_s \bra{\tau_\mathrm{L}, s}\boldsymbol{Q}(\tau_\mathrm{L}, \tau_\mathrm{R})\ket{\tau_\mathrm{R}, s},\label{eq:sw}
\end{align}
where $\boldsymbol{Q}(\tau_\mathrm{L}, \tau_\mathrm{R})$ is the product of matrices on the interval ($\tau_\mathrm{L}$, $\tau_\mathrm{R}$].
We do not have to recompute the bra and ket as long as updates are performed within the window.

We propose a few elementary updates, whose number is proportional to that of flavors, at each position of the window.
After that, we move the window to the next position with a finite overlap with the previous position as illustrated in Fig.~\ref{fig:sliding-window} (b).
We updates the bra and ket by applying creation, annihilation operators and time-evolution operators, or by loading cached data from memory.
Noted that we do not have to calculate the bra and ket from scratch thanks to the sequential move of the window on the imaginary-time axis.
The procedure is repeated by moving the window sequentially back and forth on the whole interval [$0$, $\beta$].

A reasonable value of the window width $\tau_\mathrm{win}$ is automatically estimated during the thermalization process.
We choose $\tau_\mathrm{win} = \beta/\braket{n}_\mathrm{MC}$, where $\braket{n}_\mathrm{MC}$ is the Monte Carlo average of the perturbation order. 

In the sliding-window update, we never insert/remove a pair of creation and annihilation operators which cross the boundary at $\beta \equiv 0$ as illustrated in Fig.~\ref{fig:sliding-window} (c).
To avoid this problem,
we also perform a global update in which all the operators are shifted on the imaginary-time axis by a fixed random time $\Delta \tau$ $\in$ $[0,\beta]$ [see Fig.~\ref{fig:sliding-window} (d)].
This update is proposed each time the window has completed a back-and-forth run on the whole interval.
If this update is accepted, the bra and the ket are recomputed from scratch, which costs $O(\beta)$.
However, its computational cost is typically smaller than that of the sliding-window update because it is performed less frequently.

\subsection{Restricting the trace to the active space}\label{sec:active-space}
To further reduce the computational cost,
we offer several options of eliminating high-energy states which do not contribute to low-temperature physics. 
The first one, called ``inner-outer-state cutoff", was originally introduced in Ref.~\cite{Haule:2007kx}.
Here, we simplify do not include the eigenstates of $\mathcal{H}_\mathrm{loc}$ whose energies are higher than a certain cutoff in the construction of the  block matrices.
As a consequence, the matrices appearing in Eq.~(\ref{eq:sectored-trace}) are reduced in size.

The second option is called ``outer-state cutoff",
and was originally introduced in the Krylov algorithm~\cite{Lauchli:2009er}.
In Eq.~(\ref{eq:ket}), the right most matrix is modified as
\begin{align}
 \boldsymbol{E}^s(\tau_1) &\rightarrow \boldsymbol{E}^s(\tau_1)\boldsymbol{P}^s,
\end{align}
where $\boldsymbol{P}^s$ is a projector to the active eigenstates in the sector $s$, e.g., eigenstates whose energies are lower than a certain value.
We replace the left most matrix in the bra Eq.~(\ref{eq:bra}) in a similar way as well.
If we keep only a few eigenstates in the projector, the cost will be reduced from $O(N_\mathrm{H}^3)$ to $O(N_\mathrm{H}^2)$.
This may lead to a substantial speed-up when the linear dimensions of the block matrices is larger than $\approx 20$.

In Ref.~\cite{Lauchli:2009er}, it was demonstrated that at low temperatures, the approximate result converges to the exact result even when only the ground states of $\mathcal{H}_\mathrm{loc}$ are kept in the projector.
This approximation is safer than the ``inner-outer-state cutoff" because all eigenstates remain accessible during the imaginary-time evolution.

To further reduce computational costs, we adopt the lazy-trace evaluation method~\cite{Semon:2014jz}.

\section{Single-particle basis transformation}\label{sec:rotation}
In general, the average sign $\braket{\mathrm{sign}}_Z$ depends on your choice of a single-particle basis used for expanding the partition function with respect to the hybridization function in Eq.~(\ref{eq:z-cthyb-tordered}).
It is practically useful to use a single-particle basis which gives a larger average sign to improve the statistics.

To this end, our CT-HYB solver provides a functionality which allows the user to perform the expansion using an arbitrary single-particle local basis.
A transformed single-particle basis is defined as 
\begin{eqnarray}
\tilde{c}_a &=& \sum_b U_{ba}^* c_b,\\
\tilde{c}^\dagger_a &=& \sum_b U_{ba} c^\dagger_b,
\end{eqnarray}
with $U_{ab}$ being a unitary matrix, which is an input parameter.
The local Hamiltonian and the hybridization function are transformed to the new single-particle basis as described in Ref.~\cite{Semon:2012ik,Shinaoka:2015wq} before a Monte Carlo simulation.
The user inputs $t_{ab}$, $V_{abcd}$ and $\Delta_{ab}$ in the original basis to the impurity solver: The solver takes care of their transformation.

During the Monte Carlo simulation, we measure all the observables such as the single-particle Green's function in the transformed basis.
Then, the measured results are transformed back to the original basis after the Monte Carlo simulation except for the density-density correlations.

For instance, we collect the single-particle Green's function
\begin{align}
\tilde{G}_{ij}(\tau - \tau^\prime) &= -\frac{\mathrm{Tr}[T_\tau e^{-S_\mathrm{imp}} \tilde{c}_i(\tau) \tilde{c}^\dagger_j(\tau^\prime)]}{Z},\label{eq:def--G1-rot}
\end{align}
during the Monte Carlo simulation.
Once the simulation is done,
the data are transformed back to the original basis as
\begin{align}
 G_{ij} (\tau-\tau^\prime) &=  \sum_{\tilde{i}\tilde{j}}U_{i\tilde{i}}\tilde{G}_{\tilde{i}\tilde{j}} (\tau-\tau^\prime) (U^\dagger)_{\tilde{j}j}.
\end{align}
This functionality allows the user to choose any arbitrary basis transformation to improve the statistics.

\section{Usage}
The CT-HYB code is built on an updated version of the core libraries of ALPS (Applications and Libraries for Physics Simulations libraries) [ALPSCore libraries]~\cite{Gaenko:2016ic}, the Boost libraries, and Eigen.
Eigen is a C++ template header-file-only library for linear algebra.
They must be pre-installed.
One needs a MPI C++ compiler which supports C++03 to build the solver.
At runtime, one can choose either a complex-number solver or a real-number solver.
The two solvers have exactly the same interface.
The CT-HYB solver reads parameters from a text file.
In the next section, we discuss several examples.

The latest version of the code is available from a public Git repository at \url{https://github.com/ALPSCore/CT-HYB}.
One can also find a more detailed description of usage in Wiki documentation pages at
\url{https://github.com/ALPSCore/CT-HYB/wiki}.

\section{Example: Three-orbital $t_\mathrm{2g}$ model with Slater-Kanamori interaction}
\subsection{Model}
We consider a three-orbital model for the $t_\mathrm{2g}$ shell of $d$ orbitals with a Slater-Kanamori interaction and spin-orbit coupling.
Its Hamiltonian is given by
\begin{align}
\mathcal{H}&= \sum_{ij}^3 \sum_{\sigma\sigma^\prime} h^\mathrm{LS}_{i\sigma,j\sigma^\prime} c^\dagger_{i\sigma} c_{j\sigma^\prime} \nonumber\\
& + \frac{1}{2}\sum_{ijkl} \sum_{\sigma\sigma^\prime} V_{ijkl} c^\dagger_{i\sigma} c^\dagger_{j\sigma^\prime} c_{k\sigma^\prime} c_{l\sigma},\label{eq:ex1-Himp}
\end{align}
where $c^\dagger_{i\sigma}$ and $c_{i\sigma}$ are 
creation/annihilation operators of an electron at orbital $i$ with spin $\sigma$.
The first term denotes the spin-orbit coupling for the $t_{2g}$ basis.
If the states are ordered as $d_{xy\uparrow}$, $d_{xy\downarrow}$, $d_{yz\uparrow}$, $d_{yz\downarrow}$, $d_{zx\uparrow}$, $d_{zx\downarrow}$, its 
matrix elements are 
\begin{align}
h_{i\sigma,j\sigma^\prime} &= 
\frac{\zeta}{2}
\left(
\begin{array}{cccccc}
 0   &   0   &   0   &  1  &   0   &  -i \\
 0   &   0   &  -1  &   0   &  -i  &   0  \\
 0   &  -1  &   0   &   0   &  i  &   0  \\
 1  &   0   &   0   &   0   &   0   &  -i \\
 0   &  i  &  -i  &   0   &   0   &   0  \\
 i  &   0   &   0   &  i  &   0   &   0
\end{array}
 \right),
\end{align}
where $\zeta$~($>$0) is the amplitude of the spin-orbit coupling.
The none-zero elements of the Coulomb tensor are given by $V_{iiii} = U$, $V_{ijji} = U-2J$, $V_{ijij} = J$, $V_{iijj} = J$.
The chemical potential is chosen such that the system is close to half filling: $\mu = \frac{5}{2}U-5J$.

The bath consists of three non-interacting orbitals.
The intra-bath Hamiltonian reads
\begin{align}
 \mathcal{H}_\mathrm{bath} &= \sum_{i\neq j}^3 t^\prime d^\dagger_{i\sigma} d_{j\sigma} = \sum_{ij}^3 h^\mathrm{bath}_{ij} d^\dagger_{i\sigma} d_{j\sigma},\label{eq:ex1-Hbath}
\end{align}
where $t^\prime \neq 0$ gives rise to off-diagonal elements of the hybridization function.
The hybridization term reads
\begin{align}
\mathcal{H}_\mathrm{hyb} &= \lambda \sum_i^3 (c^\dagger_{i\sigma} d_{i\sigma} + d^\dagger_{i\sigma} c_{i\sigma}),\label{eq:ex1-hyb}
\end{align}
where we take the coupling $\lambda=1$.

From Eqs.~(\ref{eq:ex1-Hbath}) and (\ref{eq:ex1-hyb}),
we obtain the hybridization function
\begin{align}
\Delta_{i\sigma j\sigma^\prime}(\tau) &=  -\frac{\lambda^2\delta_{\sigma\sigma^\prime}}{\beta} \sum_{n=-\infty}^\infty \frac{e^{-\mathrm{i}\omega_n \tau}}{i\omega_n I - h^\mathrm{bath}},
\end{align}
where $I$ is an identity matrix.

\subsection{Diagonal hybridization function}
\begin{figure}
	\centering\includegraphics[width=.5\textwidth,clip]{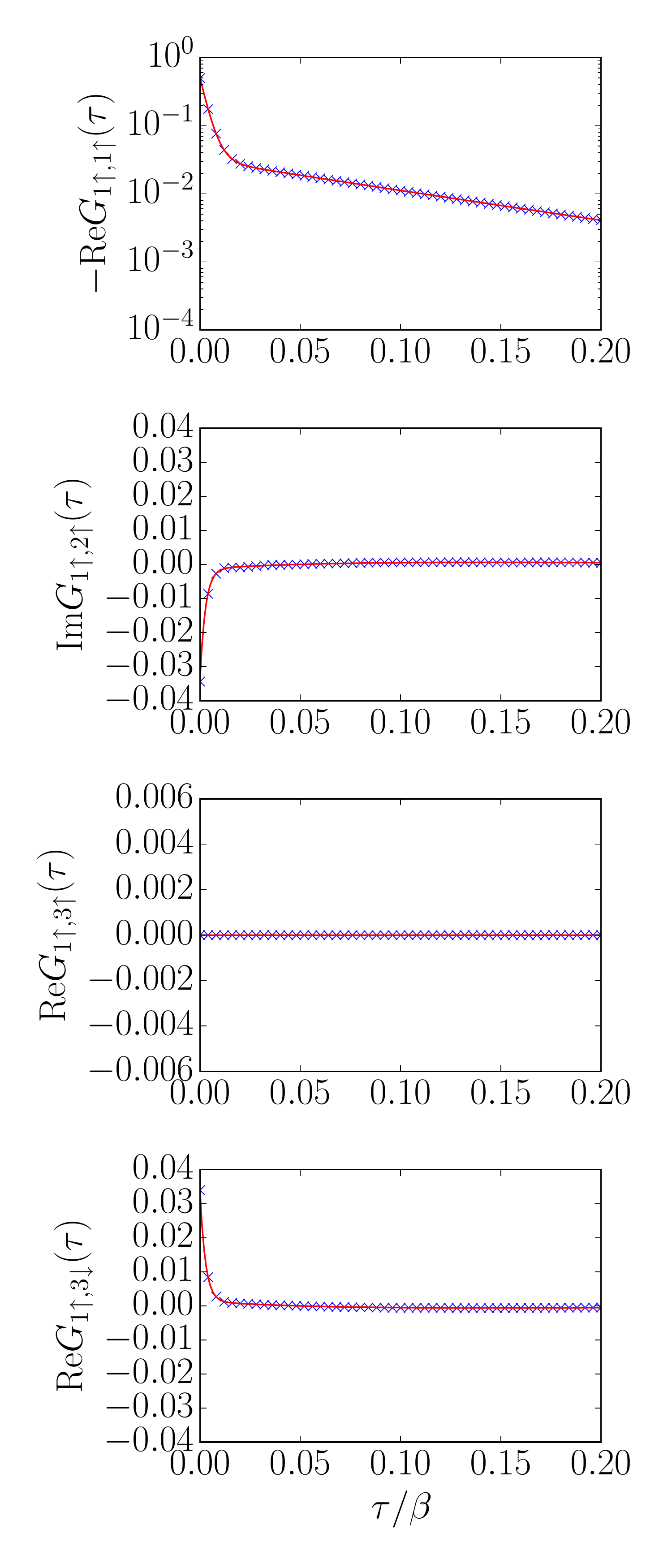}
	\caption{
		(Color online) 
		The single-particle Green's function $G(\tau)$ computed for $\zeta=1$, $U=10$, $J/U=1/4$, $\beta=40$, and $t^\prime=0$.
		The crosses show the results of the CT-HYB code, while the solid lines correspond to the exact results.
	}
	\label{fig:gf}
\end{figure}

We first solve the model with a diagonal hybridization function, i.e., for $t^\prime=0$.
We take $\zeta=1$, $U=10$, $J/U=1/4$, $\beta=40$.
In Fig.~\ref{fig:gf},
we compare the single-particle Green's function $G(\tau)$ measured by the worm sampling with numerically exact results computed by pomerol~\cite{pomerol}.
The average sign is about 0.95.
The number of Legendre polynomials is $N_l=50$.
The perturbation expansion was performed in the original $t_\mathrm{2g}$ basis.
One can see that our code can measure both the diagonal ($G_{00}$) and off-diagonal elements ($G_{02}$, $G_{05}$). 
Using the conventional sampling method, it would not be possible to measure the off-diagonal elements of $G(\tau)$ in the case of a diagonal hybridization function. 

\subsection{Off-diagonal hybridization function}
\begin{figure}
	\centering\includegraphics[width=.5\textwidth,clip]{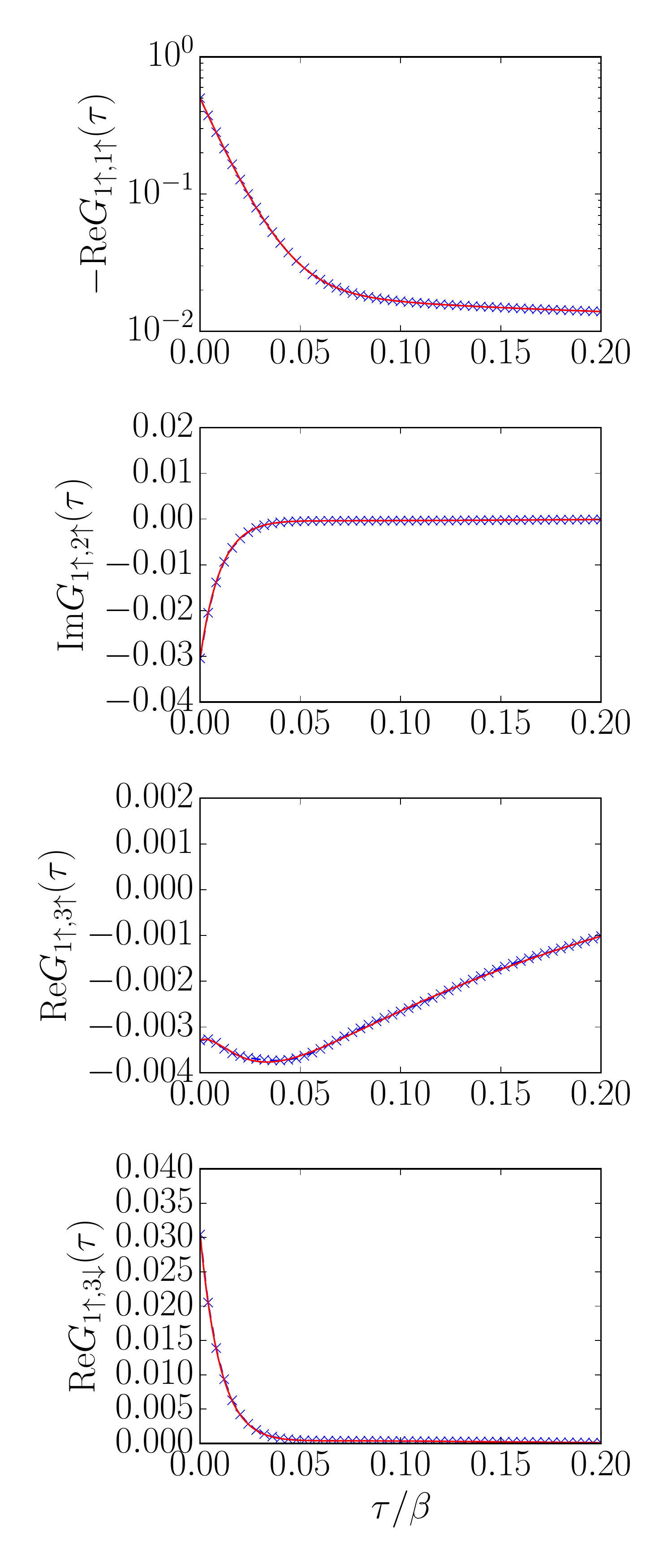}
	\caption{
		(Color online) 
		The single-particle Green's function $G(\tau)$ computed for $\zeta=1$, $U=10$, $J/U=1/4$, $\beta=10$, and $t^\prime=0.2$.
		The crosses show the results of the CT-HYB code, while the solid lines correspond to the exact results.
	}
	\label{fig:gf2}
\end{figure}

Next, we solve the model for $t^\prime=0.2$, $\zeta=1$, $U=10$, $J/U=1/4$, $\beta=10$.
In Fig.~\ref{fig:gf2},
we compare the computed results with exact results.
The number of Legendre polynomials is $N_l=30$.
The perturbation expansion was performed in the original $t_\mathrm{2g}$ basis.
The average sign is about 0.48.

\section{Summary}
We have presented an open-source C++ implementation of the continuous-time hybridization expansion Monte Carlo method for multi-orbital impurity models with general instantaneous two-body interactions and complex hybridization functions.
We have discussed the technical details of the implementation, and presented some examples of Monte Carlo simulation results for a three-orbital model, which can serve as a benchmark or reference. 

\section*{Acknowledgments}
We gratefully acknowledge support by the wider ALPS community~\cite{Bauer:2011tz,Albuquerque:2007ja}.
HS and PW acknowledge support from the Deutsche Forschungsgemeinschaft via FOR 1346, the SNSF (Swiss National Science Foundation) Grant No. 200021E-149122,
ERC Advanced Grant SIMCOFE and NCCR MARVEL.
We thank Markus Wallerberger and Florian Sohn for useful comments on the manuscript and the code.
This work was supported by JSPS KAKENHI Grant Number 15H05885 (J-Physics), 16K17735.
EG was supported by DOE ER 46932.
Part of the calculations were performed on the ISSP supercomputing system.
 
\section*{References}
\bibliographystyle{elsarticle-num}
\bibliography{ref.bib}

\appendix
\section{Estimation of volumes of configuration spaces}\label{sec:wang-landau}
We give a brief description on how to estimate the volumes of configuration spaces in the worm sampling using the Wang-Landau algorithm~\cite{Wang:2001eb,Wang:2001jk}.
Let us consider the case of measuring $N$ observables $\mathcal{O}_1, \cdots, \mathcal{O}_N$ by the worm sampling.
The corresponding configuration spaces are $C_{\mathcal{O}_1}, \cdots, C_{\mathcal{O}_N}$, while that of the partition function is $C_Z$.

We design our Monte Carlo dynamics so that we spend a roughly equal number of steps in each of the $N+1$ subspaces.
This is done by choosing $\eta_i = V_Z/V_i$ in the measurement process,
where $V_i$ is the volume of subspace $i$ ($i=\mathcal{O}_1, \cdots, \mathcal{O}_N$).

During the thermalization processes,
we estimate the subspace volumes following the standard procedure of the Wang-Landau algorithm.
In practice, we start a Monte Carlo simulation with an initial guess $V_j = 1$ ($j=Z, \cdots, \mathcal{O}_N$).
The acceptance rate of a worm insertion/removal is computed using the current values of the weights as $\eta_i = V_Z/V_i$ ($i=\mathcal{O}_1, \cdots, \mathcal{O}_N$).
After each attempt of a worm insertion or removal,
we update $V_j$ as $V_j\rightarrow \lambda V_j$, where $j$ is the current subspace and $\lambda$ ($>1$) is a modification factor.
This forces the configuration to visit all the subspaces.
During this random walk between the subspaces, we count the number of Monte Carlo steps $N_j$ spent in each subspace ($j=Z, \mathcal{O}_1, \cdots, \mathcal{O}_N$).
This random walk is performed until the histogram $\{N_j \}$ becomes reasonably \textit{flat}.
We found that a maximum deviation of 20\% from the mean value is a reasonable criterion.
Once this criterion is met, we reset the histogram to zero, and update the modification factor as $\lambda\leftarrow\sqrt{\lambda}$.
A new random walk is performed until a flat histogram $\{N_j\}$ is again obtained with the new (smaller) modification factor.
We repeat this procedure until $\{V_j\}$ ($j=Z, \mathcal{O}_1, \cdots, \mathcal{O}_N$) converges within a reasonable accuracy as $\lambda$ converges to 1.

\section{Auto partitioning of the local Hilbert space}\label{sec:partioning}
\begin{figure}
	\centering\includegraphics[width=.45\textwidth,clip]{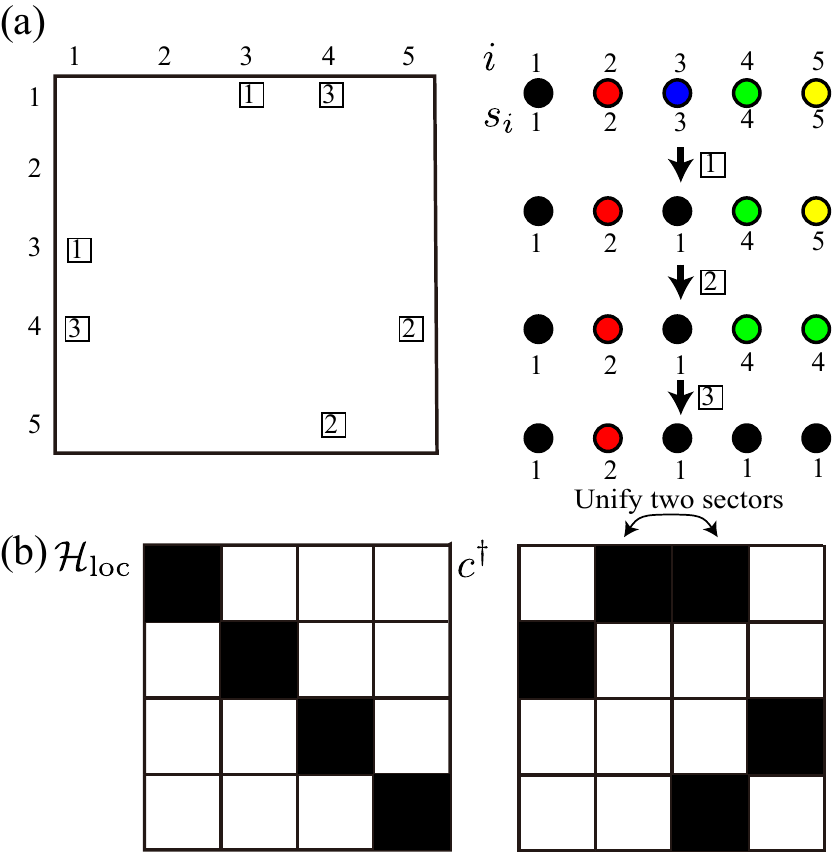}
	\caption{
		(Color online) 
		(a) Partitioning of a $5\times 5$ Hamiltonian.
		The $i$-th element in the occupation basis belongs to the sector $s_i$.
		Sectors are merged by adding off-diagonal elements to $H_\text{loc}$, one by one (squares with numbers).
		(b) Partitioning of the creation operator.
	}
	\label{fig:partitioning}
\end{figure}
We illustrate our procedure for the auto partitioning of the Hilbert space in Fig.~\ref{fig:partitioning}.
First, we compute the matrix elements of $\mathcal{H}_\mathrm{loc}$ in the occupation basis for a given single-particle basis.
In the first step [Fig.~\ref{fig:partitioning}(a)],
we partition the Hilbert space into subspaces so that $\mathcal{H}_\mathrm{loc}$ becomes a block diagonal matrix.
In the second step [Fig.~\ref{fig:partitioning}(b)],
we unify some of the sectors so that the block matrices of creation and annihilation operators have only a single nonzero block for each row.

If there is no non-zero element, i.e., $\mathcal{H}_\mathrm{loc}=0$,
each vector in the occupation basis forms its own subspace (sector).
Hereafter, we denote the sector to which the $i$-th element of the occupation basis belongs by $s_i$.
We add non-zero elements into the matrix of $\mathcal{H}_\mathrm{loc}$ one by one.
Each time we add a new nonzero element $H_{ij}\neq 0$,
we unify the two sectors $s_i$ and $s_j$, if those two vectors belong to different sectors ($s_i\neq s_j$).
The produce is illustrated in Fig.~\ref{fig:partitioning}(a) for a $5\times 5$ matrix.
After going through all the nonzero elements and reordering rows and columns appropriately,
the matrix $\mathcal{H}_\mathrm{loc}$ becomes block-diagonal.

In the second step [Fig.~\ref{fig:partitioning}(b)],
we compute the elements of the block matrices of creation operators.
If more than two blocks are nonzero in a given row, we unify the corresponding two sectors.
This procedure is repeated for each column as well.

The elementary operation of this procedure is unifying two sectors $s$ and $s^\prime$ for given $s$ and $s^\prime$ ($s < s^\prime$).
A naive procedure would be to scan through all vectors having $s^\prime$ and assign $s$ to them ($s_i\leftarrow p$ if $s_i = s^\prime$).
But, this is computationally inefficient for a large number of orbitals because we have to scan the whole Hilbert space many times during the partitioning of the matrices.
Instead, we adopt the Hoshen-Kopelman single-pass cluster identification algorithm~\cite{Hoshen:1976vg}.

\section{Measurement of single-particle Green's function by reconnecting hybridization lines}\label{sec:meas-G}
Here we describe how to generate multiple configurations for the measurement of the single-particle Green's function.
First, we present a general procedure for measuring multiple samples from a single configuration $c_0$ in the worm space $C_A$ for an observable $A$.
We assume that the thermodynamic average of $A$ is given by
\begin{align}
\braket{A} &= \braket{f(c_A)}_\mathrm{MC},
\end{align}
where $\braket{\cdots}_\mathrm{MC}$ is the Monte Carlo average, and $f$ is some function of a configuration in $C_A$.

At the measurement step,
we randomly generate a set of $N$ configurations,
$S = \{c_0, c_1, c_2, \cdots, c_{N-1}\}$,
including the current one $c_0$ ($N>1$).
$P_{c\rightarrow S}$ denotes the probability that the set $S$ is generated when the current configuration is $c$.
Here, we require $P_{c_i \rightarrow S} = P_{c_j \rightarrow S}$ for $0 \le i\neq j \le N-1$.
Then, it is easy to prove that 
\begin{align}
   \braket{A} &= \Braket{\frac{\sum_i w_i f(c_i)}{\sum_i w_i}}_\mathrm{MC}.
\end{align}

\begin{figure}
	\centering\includegraphics[width=.45\textwidth,clip]{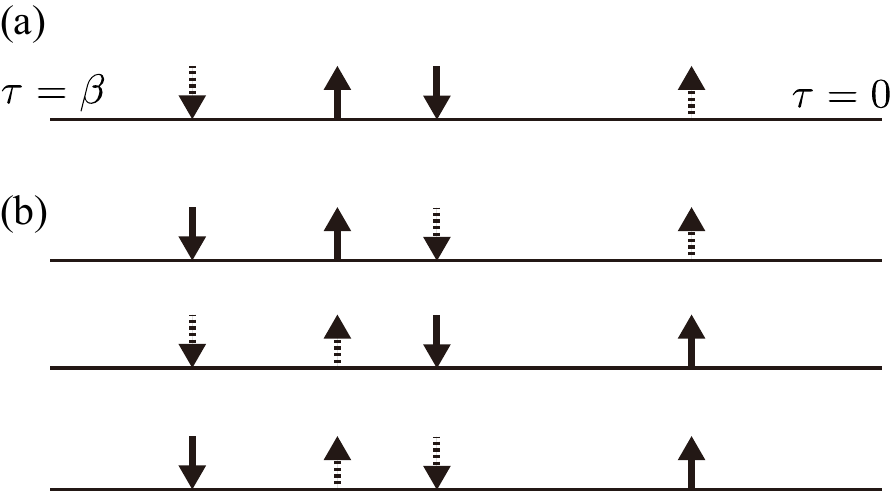}
	\caption{
		(a) Illustration of a configuration with a worm for the single-particle Green's function.
		Up and down arrows with solid lines represent annihilation and creation operators coupled with the bath, respectively.
		Annihilation and creation operators of the worm are denoted by dashed arrows.
		(b) Three configurations can be generated by reconnecting hybridization lines from the configuration in (a).
	}
	\label{fig:measure-G1}
\end{figure}
Let us consider a configuration in the $\G1$ space illustrated in Fig.~\ref{fig:measure-G1}(a).
In this case, we can generate three additional configurations shown in Fig.~\ref{fig:measure-G1}(b) by reconnecting hybridization lines,
which does not require the reevaluation of the trace.

We compute the relative Monte Carlo weights of these configurations as follows.
(i) First, we attach hybridization lines to the creation and annihilation operators of the worm.
(ii) Then, we detach a creation operator and an annihilation operator from the bath.
It should be noted that, to avoid a singular matrix, we need to introduce auxiliary fields,
which will be removed again in step (ii).

For step (i), we define the following matrix of hybridization functions,  
\begin{align}
\boldsymbol{M}^\prime &=
\left[
\begin{array}{ccc|cc}
 &                  &  & \Delta_{a^\prime_1, b} & 0\\
 & \boldsymbol{M}_0 &  & \vdots  & \vdots \\
 &                  &  & \Delta_{a^\prime_N, b} & 0\\
 \hline
 \Delta_{a, a_1}  & \cdots  & \Delta_{a, a_N} & \Delta_{a, b}     & s_1\\
 0                & \cdots  & 0               &     s_2           & s_3
\end{array}
\right],
\end{align}
where $\boldsymbol{M}_0$ is the matrix defined in Eq.~(\ref{eq:M}) for the current configuration, e.g., the one shown in Fig.~\ref{fig:measure-G1}(a).
The auxiliary fields $s_1$, $s_2$, and $s_3$ in the last column and row were introduced to avoid a singular matrix, i.e, $\mathrm{det} \boldsymbol{M}^\prime = 0$ for $\Delta_{a,b}=0$.
We compute the inverse matrix $(\boldsymbol{M}^\prime)^{-1}$ and the ratio $\mathrm{det}\boldsymbol{M}^\prime/ \mathrm{det} \boldsymbol{M}_0$ from $\boldsymbol{M}_0^{-1}$,  $\Delta_{a_i^\prime, b}$, and $\Delta_{a, a_j}$ using the fast update formula~\cite{Gull:2011jda}.
The choice of the auxiliary fields may be arbitrary a long as a singular matrix is avoided, because the last column and row will be removed.
In practice, we adopt $s_i = \pm \delta$,
with $\delta$ being a small number (typically around $10^{-5}$).

Once $(\boldsymbol{M}^\prime)^{-1}$ is computed, we are ready to compute the relative weight $w_{ij}$ of all the configurations generated by reconnecting hybridization lines,
where $i$ and $j$ represent creation and annihilation operators detached from the bath.
Counting sign changes from permutations of columns and rows,
the relative weight is obtained as 
\begin{align}
 \Delta w_{ij} &\equiv w_{ij}/w_{nn} = \left(\mathrm{det}\boldsymbol{M}^\prime/ \mathrm{det} \boldsymbol{M}_0\right) \nonumber\\
 &
 \times \mathrm{det}
 \left(
  \begin{array}{cc}
  ({\boldsymbol{M}^\prime}^{-1})_{i,j} & ({\boldsymbol{M}^\prime}^{-1})_{i,n+1}\\
  ({\boldsymbol{M}^\prime}^{-1})_{n+1,j} & ({\boldsymbol{M}^\prime}^{-1})_{n+1,n+1}
  \end{array}
  \right),\label{eq:relative-weight-G1}
\end{align}
where $n$ is the perturbation order.
Here, we removed the last column and row as well to avoid systematic errors from the auxiliary fields.
In practice, the Monte Carlo average in Eq.~(\ref{eq:G1-legendre}) is replaced by the weighted averaged over the configurations
\begin{align}
& \Braket{
\frac{
\sum_{ij}^n \Delta w_{ij}
	\mathrm{sign}(w_{nn}) \mathrm{sign}(\Delta w_{ij}) \tilde{P}_l(x(\tau_j-\tau_i^\prime))
}
{\sum_{ij}^n \mid w_{ij}/w_{nn}\mid }
}_\G1,
\end{align}
where $\mathrm{sign} (w_{nn})$ is the sign of the Monte Carlo weight of the current configuration.

\section{Measurement of the two-particle Green's function}\label{sec:meas-G2}
We measure the two-particle Green's function in the mixed basis representation proposed in Ref.~\cite{Boehnke:2011dd}:
\begin{align}
 & G^{(2)}_{abcd}(\tau_{12}, \tau_{34}, \tau_{14})=\nonumber\\
 & 
 \sum_{l,l^\prime \ge 0} \sum_{m\in \mathbb{Z}}
 \frac{\sqrt{2l+1}\sqrt{2l^\prime+1}}{\beta^3}(-1)^{l^\prime+1} \nonumber\\
&\times P_l(x(\tau_{12})) P_{l^\prime}(x(\tau_{34})) e^{\mathrm{i}\omega_m\tau_{14}} G^{(2)}_{abcd}(l,l^\prime,\mathrm{i}\omega_n),
\end{align}
where $\omega_n = 2m\pi/\beta$.

A worm for the two-particle Green's function consists of four operators $c_a(\tau_a) c^\dagger_b (\tau^\prime_b) c_c(\tau_c) c^\dagger_d(\tau_d^\prime)$ with four independent flavor and time variables.
In the mixed basis representation,
the worm estimator reads
\begin{align}
 & G^{(2)}_{abcd}(l,l^\prime,i\omega_n) = \mathcal{N}
 \frac{\sqrt{2l + 1}\sqrt{2l^\prime+ 1}}{\beta} (-1)^{l^\prime + 1} \nonumber\\
 & \times\langle (M_{ab}M_{cd} - M_{ad}M_{cb}) \nonumber \\
 & \tilde{P}_l(\tau_a-\tau_b^\prime)\tilde{P}^\prime_l(\tau_c-\tau_d^\prime)e^{i\omega_n (\tau_a - \tau_d^\prime)}\rangle_{G^{(2)}}.
\end{align}
Here, the normalization factor is given by
\begin{align}
\mathcal{N} &= \frac{N_{G^{(2)}} }{\eta_{G^{(2)}} \braket{\mathrm{sign}} N_Z},
\end{align}
where $N_{G^{(2)}}$ and $N_Z$ represent the numbers of Monte Carlo steps in the worm space and the partition function space, respectively.

As for the single-particle Green's function, we generate multiple samples following the procedure described below.
First, we attach hybridization lines to the creation and annihilation operators of the worm by adding three columns and rows 
-- one of each made up of auxiliary fields -- to the hybridization-function matrix. 
Then, we remove three columns and rows.

We define the matrix of the intermediate state by
\begin{align}
\boldsymbol{M}^\prime &=
\left[
\begin{array}{ccc|ccc}
&                  &  & \Delta_{a_1, b} & \Delta_{a_1, d} & 0\\
& \boldsymbol{M}_0 &  & \vdots  & \vdots & \vdots \\
&                  &  &\Delta_{a_N, b} & \Delta_{a_1, d} & 0 \\
\hline
\Delta_{a, a_1^\prime}  & \cdots  & \Delta_{a, a_N^\prime} & \Delta_{a, b}  & \Delta_{a, d} & s_1\\
\Delta_{c, a_1^\prime}  & \cdots  & \Delta_{c, a_N^\prime} & \Delta_{c, b}  & \Delta_{c, d} & s_2\\
0                & \cdots  & 0             &     s_3        & s_4           & s_5
\end{array}
\right],
\end{align}
where $s_i$ ($i = 1, \cdots, 5$) are auxiliary fields.
Now, we define $a_{N+1}^\prime \equiv b$, $a_{N+2}^\prime \equiv d$, $a_{N+1} \equiv a$, and $a_{N+2} \equiv c$.
Similarly to Eq.~(\ref{eq:relative-weight-G1}),
the relative weight of the worm 
$c_{a_\alpha} c^\dagger_{a^\prime_\beta} c_{a_\gamma} c^\dagger_{a^\prime_\eta}$
is given by
\begin{align}
\Delta w_{\alpha\beta\gamma\eta} =& \left(\mathrm{det}\boldsymbol{M}^\prime/ \mathrm{det} \boldsymbol{M}_0\right) \nonumber\\
& \times \mathrm{det} ({\boldsymbol{M}^\prime}^{-1}; \alpha,\gamma,n+1: \beta,\eta,n+1),\label{eq:relative-weight-G2}
\end{align}
where we use the shorthand notation
\begin{align}
\mathrm{det}(\boldsymbol{A}; i_1 i_2 i_3 : j_1 j_2 j_3) &= \mathrm{det}
  \left(
  \begin{array}{ccc}
  A_{i_1, j_1} &  A_{i_1, j_2} &  A_{i_1, j_3}\\
  A_{i_2, j_1} &  A_{i_2, j_2} &  A_{i_2, j_3}\\
  A_{i_3, j_1} &  A_{i_3, j_2} &  A_{i_3, j_3}\\
  \end{array}
  \right).
\end{align}
We omitted the index for imaginary time to simplify the notation.
In contrast to the measurement of the single-particle Green's function,
we do not take the summation of $\alpha$, $\beta$, $\gamma$, and $\eta$ over all columns and rows,
because it would cost $O(n^4)$, which is more expensive than the Monte Carlo updates [$O(n^3)$]. 
Instead, we generate two sets $S_\mathrm{col}$ and $S_\mathrm{row}$ by selecting a few elements from 1, 2, $\cdots$, $n+1$, $n+2$ so that they always include $n+1$ and $n+2$.
The size of the two sets $n^\prime$ is taken to be typically around 5--10.
The summation is taken over these two sets as $\alpha, \gamma \in S_\mathrm{col}$ and  $\beta, \eta \in S_\mathrm{row}$.

\end{document}